\documentclass[floats,twocolumn,aps,prb,longbibliography,superscriptaddress]{revtex4-2}

\usepackage{graphicx}
\usepackage{amsfonts}
\usepackage{amssymb}
\usepackage{amsbsy}
\usepackage{amsmath}\usepackage[colorlinks = true,
            linkcolor = magenta,
            urlcolor  = magenta,
            citecolor = magenta,
            anchorcolor = magenta]{hyperref}
\usepackage{hyperref}
\usepackage{xcolor}
\usepackage{subeqnarray}
\usepackage{cleveref}
\usepackage{tabularx}
\usepackage{tabularray}
\allowdisplaybreaks[1]

\newcommand*{\balancecolsandclearpage}{%
  \close@column@grid
  \cleardoublepage
  \twocolumngrid
}

\def\P{\mathcal P}

\begin{document}

\title{Finite-temperature phase transitions in $S=1/2$ three-dimensional Heisenberg magnets from high-temperature series expansions}

\author{M. G. Gonzalez}
\affiliation{Sorbonne Universit\'e, CNRS, Laboratoire de Physique Th\'eorique de la Mati\`ere Condens\'ee, LPTMC, F-75005 Paris, France}
\affiliation{Helmholtz-Zentrum Berlin f\"ur Materialien und Energie, Hahn-Meitner Platz 1, 14109 Berlin, Germany}
\affiliation{Dahlem Center for Complex Quantum Systems and Fachbereich Physik, Freie Universität Berlin, 14195 Berlin, Germany\looseness=-1}

\author{B. Bernu}
\affiliation{Sorbonne Universit\'e, CNRS, Laboratoire de Physique Th\'eorique de la Mati\`ere Condens\'ee, LPTMC, F-75005 Paris, France}

\author{L. Pierre}
\affiliation{Université Paris X, UFR SGMI (Sciences gestion, mathématique et informatique), F-92000 Nanterre, France\looseness=-1}

\author{L. Messio}
\affiliation{Sorbonne Universit\'e, CNRS, Laboratoire de Physique Th\'eorique de la Mati\`ere Condens\'ee, LPTMC, F-75005 Paris, France}
\affiliation{Institut Universitaire de France (IUF), F-75005 Paris, France}

\begin{abstract} 
Many frustrated spin models on three-dimensional (3D) lattices are currently being investigated, both experimentally and theoretically, and develop new types of long-range orders in their respective phase diagrams. They present finite-temperature phase transitions, most likely in the Heisenberg 3D universality class. However, the combination between the 3D character and frustration makes them hard to study. We present here several methods derived from high-temperature series expansions (HTSEs), which give exact coefficients directly in the thermodynamic limit up to a certain order; for several 3D lattices, supplementary orders than in previous literature are reported for the HTSEs. We introduce an interpolation method able to describe thermodynamic quantities at $T > T_c$, which we use here to reconstruct the magnetic susceptibility and the specific heat and to extract universal and non-universal quantities (for example critical exponents, temperature, energy, entropy, and other parameters related to the phase transition). While the susceptibility associated with the order parameter is not usually known for more exotic long-range orders, the specific heat is indicative of a phase transition for any kind of symmetry breaking. We present examples of applications on ferromagnetic and antiferromagnetic models on various 3D lattices and benchmark our results whenever possible.
\end{abstract}

\date{\today}

\maketitle 

\section{Introduction}
\vspace{-0.3cm}

The quantum Heisenberg model was first introduced to explain why certain compounds developed a spontaneous magnetization when cooled below a given temperature (called the critical temperature), even in the absence of an applied magnetic field. Its success in explaining these so-called ferromagnets placed the model at a central spot in the study of quantum magnetism. Since then, a wide variety of three-dimensional (3D) compounds have been investigated, whose interactions can be mostly described by ferro- or antiferromagnetic Heisenberg interactions. More often than not, these systems present finite-temperature phase transition to long-range magnetically ordered phases. Even in 2D systems, where the Mermin-Wagner theorem \cite{Mermin66} precludes the existence of magnetic order at finite temperatures, materials often present finite-temperature phase transitions due to remaining weak 3D correlations \cite{Sengupta03, Yasuda05, Tsirlin12, Tsirlin13}. Many of these phase transitions belong to the 3D Heisenberg universality class, defined by the symmetry breaking $SU(2) \to U(1)$.

Phase transitions are characterized by the critical temperature $T_c$ and exponents of the singularities present in the thermodynamic functions. Obtaining them is thus an important task, and to do so many different methods have been developed. In particular, in the ferromagnetic case, high-temperature series expansions (HTSEs) methods such as the Dlog Pad\'e or ratio methods are known to obtain numerically accurate critical temperatures for simple lattices such as the simple cubic (\textit{sc}), body-centered cubic (\textit{bcc}) and face-centered cubic (\textit{fcc}) \cite{Baker67, Oitmaa96, Oitmaa04, Oitmaa06, Kuzmin19}. More recently the list was expanded with the pyrochlore \cite{Lohmann14}, the diamond \cite{Oitmaa18, Kuzmin19}, and the semi-simple cubic (\textit{ssc}) \cite{Kuzmin20} lattices. However, for the critical exponents, the results are not so precise.

For example, the critical exponent $\gamma$ of the magnetic susceptibility $\chi(T)$ has been calculated using field theory's renormalization group on the $N$-vector model (that is in the same universality class for $N=3$), yielding $\gamma = 1.3895(50)$ \cite{Leguillou80, Guida98}. On the other hand, HTSEs calculations in the quantum spin-$\frac12$ case show higher values, $\gamma = 1.42(1)$  for the ferro- and $\gamma = 1.43(1)$ for antiferromagnetic cases \cite{Oitmaa96, Oitmaa04, Oitmaa06}. Another example of a critical exponent is the less studied $\alpha$ from the specific heat $c_v(T)$. This exponent is negative, which implies a non-divergent singularity. Instead, $c_v(T)$ presents a cusp-like behavior that reaches a maximum value with an infinite slope. To the best of our knowledge, the standard HTSEs Dlog Pad\'e and ratio methods have never been used on $c_v$ in the literature. However, indirect HTSEs calculations through scaling relations give $\alpha$ between $-0.125$ and $-0.200(15)$ \cite{Baker67, Adler93, Kuzmin19}. On the other hand, the field theory result is $\alpha = -0.122(10)$ \cite{Guida98}, showing a larger discrepancy than in the case of $\gamma$. On the experimental side, there are empirical fits on specific heat measurements that give negative values of $\alpha$ down to $-0.3$, while most of them lie around the field theory result \cite{Kaul85, Berry23}.

Finally, it would be desirable to have reliable results not only on the critical quantities (critical temperature $T_c$ and exponents) but on the thermodynamic functions at all temperatures ($c_v(T)$, $\chi(T)$). In this sense, the quantum Monte Carlo (QMC) calculations obtain reliable results, but only on finite lattices, and only in the absence of frustration due to the sign problem. Methods based on exact diagonalization and tensor network algorithms can be used on frustrated systems, but only in dimensions 1 and 2 \cite{Prelovsek18, Prelovsek20, Prelovsek20b, Gauthe22}. Other methods work directly in the thermodynamic limit like the rotationally invariant Green’s function method which obtains qualitatively good results in 3D \cite{Juhasz09, Muller15}. The pseudo-Majorana functional renormalization group provides quantitatively good results down to moderate temperatures but becomes uncontrolled at low temperatures \cite{Niggemann21, Niggemann22, Niggemann23}. Finally, the HTSEs are quasi-exact at high temperatures but fail close to the transition temperature, even when using Pad\'e approximants. An interpolation scheme of HTSEs solving this was proposed in our previous article in cases where $c_v$ presents a logarithmic divergence, such as in the 2D-Ising or 2D-XXZ models \cite{Gonzalez21}, resulting both in an evaluation of critical quantities and of $c_v(T)$ for temperatures from infinite down to $T_c$. 

In this article, we first revisit the HTSEs Dlog Pad\'e method for Heisenberg ferromagnets on several 3D lattices such as the \textit{fcc}, \textit{bcc}, \textit{sc}, pyrochlore and \textit{ssc} lattices. For most, we calculated higher orders in the HTSEs than in previous articles, using an optimized algorithm. Also, we extend the Dlog Pad\'e method to obtain quantities such as the critical energy $e_c$, the $c_v$ critical exponent $\alpha$, and non-universal quantities $A$ and $B$ ($c_v(\beta)\sim B - A (\beta_c - \beta)^{-\alpha}$). Finally, we extend the previously mentioned interpolation method \cite{Gonzalez21} to cases where $c_v$ presents a cusp-like behavior with a negative exponent $\alpha > -1 $ and to cases where $\chi$ presents a divergent singularity with positive exponent $\gamma > 0$. For some 3D lattices, we can extrapolate $c_v$ and $\chi$ at all temperatures down to $T_c$.

The remaining of the article is organized in the following way. In Sec. \ref{secMandM}, we present the HTSEs methods to study finite-temperature phase transitions, both the Dlog Pad\'e and our interpolation method. In Sec. \ref{secRes} we show our results, first for the Dlog Pad\'e, and next for the interpolation methods. Finally, conclusions and perspectives are given in Sec. \ref{secCon}.

\vspace{-0.3cm}
\section{Model and Methods}
\label{secMandM}
\vspace{-0.3cm}

The Heisenberg model is defined as
\begin{equation}
\mathcal{H} = J\sum_{\langle i j \rangle} \mathbf{S}_i \cdot \mathbf{S}_j,
\label{eq01}
\end{equation}
where $J$ is the exchange interaction, the sum $\langle ij \rangle$ runs over nearest neighbors on a 3D lattice, and $\mathbf{S}_i$ are the quantum spin-$\frac12$ operators. The classical approximation consists in replacing the operators $\mathbf{S}_i$ by 3D vectors. In the ferromagnetic case ($J <0$), the quantum ground-state energy per site $e_0$ is exactly the same as the classical one, namely
\begin{equation}
e_0 = -\frac{Z}{2} S^2
\end{equation}
where $Z$ is the coordination number of the lattice. 
Even though $e_0$ does not change when taking into account quantum fluctuations, the critical temperature $T_c$ does \cite{Oitmaa04}. On the other hand, in the antiferromagnetic case ($J > 0$) on a bipartite lattice, $e_0$ and $T_c$ are the same as in the ferromagnetic case at the classical limit, but they both change in the quantum model and $e_0$ is no longer known exactly.

To study these kinds of finite temperature phase transitions we use HTSEs. HTSEs allow to perform a series expansion of certain thermodynamic functions around $\beta = 0$, where $\beta$ is the inverse temperature ($\beta = 1/T$). Two important functions are the free energy per site $f$ and the ferromagnetic zero-field susceptibility per site $\chi$ \cite{Oitmaa96}. Their HTSEs are written:
\begin{subeqnarray}
\beta f &=& -\ln 2 - \frac{1}{n_u}\sum_{i=1}^n \frac{a_i}{4^i i!} K^i + O(K^{n+1´})
\\
\overline{\chi} &=& T\chi = \frac{1}{4} + \frac{1}{2n_u} \sum_{i=1}^m \frac{b_i}{4^i i!} K^i + O(K^{m+1}),
\label{eq-defhtse}
\end{subeqnarray}
where $a_i$ and $b_i$ are integers, $K=\beta J$, and $n_u$ is the number of spins in the unit cell. These HTSEs are typically known up to orders 13 to 15 for 3D lattices (see Table \ref{tab:HTSEorders} for the order depending on the lattice). 

The thermodynamic functions present singularities at the critical temperature. However, several methods can be used to extract information about the critical point from the first coefficients of the series. We present in Sec.~\ref{sec:standard} the most commonly used: the Dlog Pad\'e method, and pursue in Sec.~\ref{sec:interpolation} with the description of a new interpolation method.

\vspace{-0.3cm}
\subsection{Dlog Pad\'e method}
\label{sec:standard}
\vspace{-0.3cm}

We assume a thermodynamic function $f(x)$ that has a power law singularity at $x_c$, of type:
\begin{equation}
f^s(x) = A\left(x_c - x \right)^{-\theta}, 
\label{eq-sing}
\end{equation}
such that $f(x)-f^s(x)$ is analytic from $x=0$ to some $x>x_c$. $x_c$ is the critical point and $\theta$ is the critical exponent. Then, the critical point and exponent can be obtained from the Dlog Pad\'e method: the logarithmic derivative
\begin{equation}
D\ln f^s(x) = \frac{{f^s}'(x)}{f^s(x)} = \frac{\theta}{x_c - x}
\end{equation}
has a simple pole given by $x_c$, whose residue is the critical exponent $\theta$. In practice, the critical point and exponent are determined from the poles and residues of the Pad\'e approximants of the HTSE of $D\ln f(x)$.

This method has been used with $f(x) = \overline{\chi}(\beta)$ to obtain results for $\beta_c$ and the critical exponent $\gamma$ on most of the typical 3D lattices \cite{Oitmaa04,Kuzmin19,Oitmaa96}. The Dlog Pad\'e method presents a fast convergence of $\beta_c$ with the HTSE order, giving several significant digits. Notably, to the best of our knowledge, this method has never been used with other thermodynamic functions such as $c_v$ to determine $\alpha$, or to obtain the critical values of the energy $e_c$ and of the entropy $s_c$, what is now done in Sec.~\ref{sec:Dlogres_cv} and \ref{sec:Dlogres_chie}.

\vspace{-0.3cm}
\subsection{Interpolation method for cusp singularities}
\label{sec:interpolation}
\vspace{-0.3cm}

Now we propose an alternative method to extract information on the critical point, using the specific heat. This is an extension from our previously introduced interpolation method for the case of logarithmic singularities \cite{Gonzalez21}. In the present case (3D Heisenberg universality class), the singular behavior of the specific heat is expressed as
\begin{equation}
c_v^s(\beta) = - A \left(\beta_c - \beta \right)^{-\alpha}
\label{eq:cvsing}
\end{equation}
where $A$ is positive. Also, $-1 < \alpha <0$ so that there is no divergence at the critical point. Instead, $c_v$ reaches a maximum value $B$ with an infinite slope (from higher temperatures). Other confluent terms, leading to more accurate results in the classical case \cite{Adler93}, are less important in the quantum case \cite{Oitmaa96}.

We build a regular function $R(\beta)$ by removing the singular behavior from the specific heat. We explore two different ways of doing this. The first one, called the \textit{interpolation method 1} (IM1), is analogous to that of Ref.~\cite{Gonzalez21}:
\begin{equation}
R(\beta) = c_v(\beta) - c_v^s(\beta)
\end{equation}
With this definition, the value $B$ of $c_v$ at the singularity is $R(\beta_c)$. The $c_v$-HTSE coefficients are calculated for a specific model up to an order $n$, and the series of $c_v^s$ are known at all orders supposing that $\beta_c$, $A$ and $\alpha$ are known. Thus, the $R$-HTSE are obtained at order $n$, whose coefficients depend on $\beta_c$, $A$ and $\alpha$. Compared to the case with a logarithmic divergency \cite{Gonzalez21}, the parameter space has one more dimension.

The other alternative to build the regular function $R$, called \textit{interpolation method 2} (IM2) is:
\begin{equation}
R(\beta) = \frac{1}{A} \frac{ c_v^s(\beta) }{c_v(\beta) -B}
\end{equation}
Again, the $R$-HTSE can be obtained up to order $n$ and this time, it depends on $\beta_c$, $B$ (instead of $A$), and $\alpha$. Defined this way, $R(\beta_c)=1/A$. It is worth mentioning that other similar methods could be developed with other regular functions $R$. But the two proposed here are sufficiently different so if both methods give similar results, we consider the results as trustworthy.

The idea behind these kinds of methods is that the parameters to obtain $R(\beta)$ have to be well-chosen for $R$ to be a truly regular function. This means that the singularity has to be canceled exactly. When this is done, the Pad\'e approximants of $R(\beta)$ will coincide down to the critical temperature (and a little further below). The quality of a given set of parameters $\{\beta_c, A, \alpha\}$ (for IM1) or $\{\beta_c, B, \alpha\}$ (for IM2) is measured by the quality function already introduced in Ref.~\cite{Gonzalez21},
\begin{equation}
\label{eq:Q}
	Q^2 = \frac{2}{(n-1)n} \sum_{i=1}^{N_\P} \sum_{j=1}^{i-1} M_\epsilon\left(\frac{\P_i(\beta_m)-\P_j(\beta_m)}{\overline{F}(\beta_c)}\right)
\end{equation}
where $N_\P$ is the number of Pad\'e approximants $\P_i$ without singularities in the range $[0,\beta_m]$, $\beta_m$ is chosen larger that $\beta_c$ to check the regular character of $R$ beyond the critical point. We take $\beta_m = (1+\delta) \beta_c$ with $\delta = 0.05$. $M_\epsilon(x)$ is a smooth function whose value is $\lesssim 1$ when $x\ll\epsilon$, and $\gtrsim 0$ when $x\gg \epsilon$. We use  $M_\epsilon(x) = 1/(1+(x/\epsilon)^8)$ with $\epsilon = 0.005$. Finally, $\overline{F}(\beta_c)=\frac12(\P_i(\beta_c)+\P_j(\beta_c))$ is the average of the two Pad\'e approximants at $\beta_c$. This $Q$ function represents roughly the proportion of coinciding Pad\'e approximants down to the critical temperature. Parameters with $Q>0.5$ are considered as good.

Once a high-quality set of parameters is found, $c_v(\beta)$ can be reconstructed by replacing the regular function with any of its coinciding Pad\'e approximants $\P_i$, 
\begin{subeqnarray}
c_v (\beta) &=& \P_i(\beta) - A \left(\beta_c - \beta \right)^{-\alpha} \qquad \textrm{for IM1},
\\
c_v (\beta) &=& B - \frac{ \left(\beta_c - \beta \right)^{-\alpha} }{\P_i(\beta)} \qquad\qquad\,\, \textrm{for IM2}.
\end{subeqnarray}

\vspace{-0.3cm}
\subsection{Interpolation method for divergent singularities}
\label{sec:interpolation2}
\vspace{-0.3cm}

The two methods presented in the previous subsection can be extended to quantities with divergent singularities, such as the magnetic susceptibility, whose singular part writes:
\begin{equation}
\overline{\chi}^s(\beta) = C \left(\beta_c - \beta \right)^{-\gamma} 
\end{equation}
where $C$ and $\gamma$ are positive. From this point, the two methods IM1 and IM2 can be applied as in the previous subsection with the simplification that no constant term has to be taken into account (the term $B$ of the previous section can be discarded as the divergency dominates it). This leads to an important difference between the extensions of IM1 and IM2 to divergent singularities. For IM1, $C$ has to be taken into account and the parameter space consists in $\{\beta_c, C, \gamma\}$. But for IM2,
\begin{equation}
R(\beta) = \frac{1}{C} \frac{ \overline{\chi}^s(\beta) }{\overline{\chi}(\beta)} 
= \frac{ \left(\beta_c - \beta \right)^{-\gamma}  }{\overline{\chi}(\beta)},
\end{equation}
the parameter space is reduced to $\{\beta_c, \gamma\}$. Thus, we will only use IM2 to interpolate $\chi$ in the following. From this regular function, the rest of the method is the same as described in the previous subsection, and the susceptibility can be reconstructed from any of its coinciding Pad\'e approximants,
\begin{equation}
\chi(\beta) = \beta \frac{ \left(\beta_c - \beta \right)^{-\gamma} }{\P_i(\beta)}.
\end{equation}

\vspace{-0.3cm}
\section{Results}
\vspace{-0.3cm}

\label{secRes}

We have numerically but exactly calculated the HTSEs of $\beta f(\beta)$ and ${\overline \chi}(\beta)$ for several 3D lattices: the \textit{fcc}, \textit{bcc}, \textit{sc}, \textit{ssc}, and pyrochlore lattices. The maximum order $n$ depends on the lattice according to Table ~\ref{tab:HTSEorders}, where we get the same order for both $\beta f(\beta)$ and ${\overline \chi}(\beta)$. Using an improved algorithm~\cite{Pierre23}, we are able to calculate several orders more than previous works for different lattices~\cite{Oitmaa96, Kuzmin19, Kuzmin20, Derzhko20}. The new terms in the HTSEs are provided in Appendix~\ref{ap1}. 

\begin{table}
    \begin{tblr}{lccc}
        \hline
        \hline
        Lattice & this article & $n_{\beta f}$ & $n_{\overline \chi}$ \\
        \hline
        \textit{fcc} & 13 & 12 \cite{Oitmaa96} & 14 \cite{Kuzmin19} \\
        \textit{bcc} & 15 & 14 \cite{Oitmaa96} & 14 \cite{Oitmaa96} \\
        \textit{sc}  & 17 & 14 \cite{Oitmaa96} & 14 \cite{Oitmaa96} \\
        \textit{ssc} & 20 &  -  & 14 \cite{Kuzmin20}\\
        pyrochlore & 16 & 13 \cite{Derzhko20} & 12 \cite{Derzhko20} \\
        \textit{sc}-\textit{ssc}, Eq.~\eqref{eq-sc-ssc} & 13 & - & -\\
        \hline
        \hline
    \end{tblr}
\caption{The order $n$ of the HTSEs used for $\beta f$ and ${\overline \chi}$ in this article (we use the same order for both quantities), compared to orders from other articles. The lattices are \textit{fcc} (face-centered cubic), \textit{bcc} (bond-centered cubic), \textit{sc} (simple cubic), \textit{ssc} (semi-simple cubic), and the pyrochlore lattice. \textit{sc-ssc} is a model interpolating between \textit{sc} and \textit{ssc}, defined in the main text. The new orders are provided in Appendix~\ref{ap1}}
\label{tab:HTSEorders}
\end{table}

In what follows, we study mostly the ferromagnetic Heisenberg model ($J=-1$ in Eq.~\eqref{eq01}) on said lattices. However, all the methods dependent on $\beta f(\beta)$ can be directly applied to the antiferromagnetic models by transforming $J$ accordingly. This means that no new HTSEs need to be calculated. We show this at the end of Sec.~\ref{subseccv} for the \textit{bcc} and \textit{sc} lattice. On the other hand, in the antiferromagnetic case, ${\overline \chi}(\beta)$ only presents a weak singularity if there is a finite-temperature phase transition. Because of this, it is usually better to use the HTSEs of the susceptibility associated with the magnetic order, which has to be calculated for each lattice. Except for the \textit{fcc} and pyrochlore lattices, all the lattices mentioned above are bipartite, and thus the corresponding susceptibility is the susceptibility associated to a staggered magnetic field \cite{Oitmaa04}. For the \textit{fcc} and pyrochlore lattices, or any other non-bipartite lattice, the existence and nature of a phase transition in the antiferromagnetic case is not trivial \cite{Schick20, Schick22, Schafer20, Astrakhantsev21, Hering22}; and therefore the definition of the susceptibility related to the order parameter is more complicated. It should also be mentioned that only ${\overline \chi}(\beta)$ associated to an uniform magnetic field is accessible experimentally.

\vspace{-0.3cm}
\subsection{Dlog Pad\'e method applied to $\overline{\chi}(\beta)$}
\label{sec:Dlogres_chi}
\vspace{-0.3cm}

\begin{figure}[!t]
\begin{center}
\includegraphics*[width=0.4\textwidth]{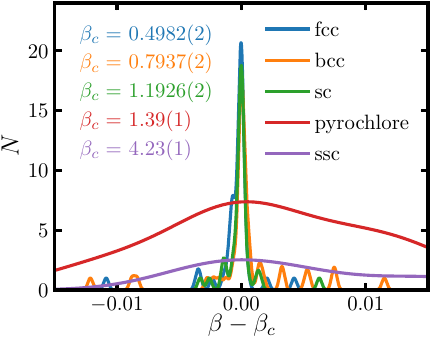}
\caption{Density of poles as defined in Eq.~\ref{eqN} from the Dlog Pad\'e method on $\overline{\chi}(\beta)$ for the ferromagnetic cases on the \textit{fcc}, \textit{bcc}, \textit{sc}, \textit{ssc} and pyrochlore lattices as a function of $\beta - \beta_c$. For each curve, the poles from the 4 highest orders are taken.}
\vspace{-0.5cm}
\label{fig1}
\end{center}
\end{figure}

We use first the Dlog Pad\'e method on $\overline{\chi}(\beta)$, which is the standard method to obtain the values of $T_c$ and the critical exponent $\gamma$ from HTSEs \cite{Oitmaa96, Oitmaa06}. The results are shown in Fig. \ref{fig1} for the ferromagnetic Heisenberg model on the \textit{fcc}, \textit{bcc}, \textit{sc}, pyrochlore, and \textit{ssc} lattices. Taking into account all the poles $\beta_i$ of the Pad\'e approximants of the logarithmic derivative of $\overline{\chi}(\beta)$, we define the density of poles as a sum of Gaussian distributions:
\begin{equation}
N(\beta) = \sum_i e^{-\frac{1}{2}\left(\frac{\beta_i-\beta}{\sigma}\right)^2}
\label{eqN}
\end{equation}
where $\sigma = 0.0002$ for the first three lattices and $\sigma = 0.005$ for the latter. For each lattice, we use the poles from the four highest orders in the corresponding HTSE.

The \textit{fcc} lattice has the highest coordination number $Z=12$, and the highest critical temperature (smallest $\beta_c$) of all lattices studied. As a consequence, the HTSE exploitation leads to high-quality results, even though the large $Z$ limits the highest order $n$ that can be reached. We can see from Fig.~\ref{fig1} that the values of $\beta_c$ are concentrated around a well-defined value, $\beta_c = 0.4982(2)$, in agreement with the previous calculations with the same method \cite{Oitmaa96, Kuzmin19}. For the \textit{bcc} lattice ($Z=8$) we get $\beta_c = 0.7937(2)$, and for the \textit{sc} lattice $\beta_c = 1.1926(2)$, both of which agree with previous results~\cite{Oitmaa96}. For the pyrochlore lattice we obtain $\beta_c = 1.39(1)$. This last value is larger than in the \textit{sc} lattice even though they both have the same coordination number. It has been argued that this is caused by the contribution of antiferromagnetic states to the partition function which is more important in the pyrochlore than in the \textit{sc} lattice \cite{Schmalfuss05, Lohmann14}. Due to frustration, the energy difference between ferro- and antiferromagnetic states in the pyrochlore lattice is less than in the \textit{sc} lattice. Finally, for the \textit{ssc} lattice, the poles are too scarce and scattered to extract accurate values of $\beta_c$ (lower peaks in Fig.~\ref{fig1} while using a higher $\sigma$). This is not surprising in the case of the \textit{ssc}, even at orders as high as 20: because of the low coordination number $Z=3$, $\beta_c$ is large and the system is close to the limit $Z=2$ (where the system can be mapped into a 1D chain, with no singularity at finite temperatures). 

\begin{figure}[!t]
\begin{center}
\includegraphics*[width=0.4\textwidth]{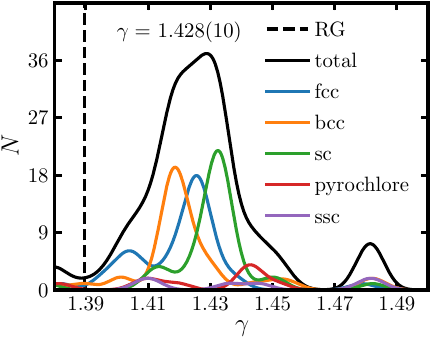}
\caption{Density of residues ($\gamma$) from the Dlog Pad\'e method on $\overline{\chi}(\beta)$ for the ferromagnetic case on the \textit{fcc}, \textit{bcc}, \textit{sc}, \textit{ssc} and pyrochlore lattices as a function of $\gamma$. For each lattice, we use the 4 highest orders. The dashed line indicates the result from field theory renormalization group \cite{Leguillou80,Guida98}.
} 
\vspace{-0.5cm}
\label{fig2}
\end{center}
\end{figure}

From the residues, we can calculate the value of the critical exponent $\gamma$. Since all five lattices belong to the same universality class, $\gamma$ is the same and thus we gather all the results in Fig. \ref{fig2}. In this case, we use Eq.~\eqref{eqN} with $\beta_i\to\gamma_i$ and $\sigma = 0.004$. As can be seen in the figure, the residues from the pyrochlore and \textit{ssc} lattices do not contribute significantly to the final result. In total we get $\gamma = 1.428(10)$, in agreement with previous results \cite{Oitmaa96, Kuzmin19}. This is different from the renormalization group value, $\gamma = 1.3895(50)$ \cite{Leguillou80, Guida98}, see dashed line in Fig. \ref{fig2}. It was proposed that this discrepancy comes from the low order of the HTSE and that higher orders might bring the numbers closer together, as it happens for the Ising model \cite{Guttmann87, Oitmaa96, Oitmaa06}. However, the present inclusion of higher orders does not seem to point in that direction. This is indicating that a lot more orders would be needed to see an appreciable shift toward the renormalization group value.

\begin{figure}[!t]
\begin{center}
\includegraphics*[width=0.4\textwidth]{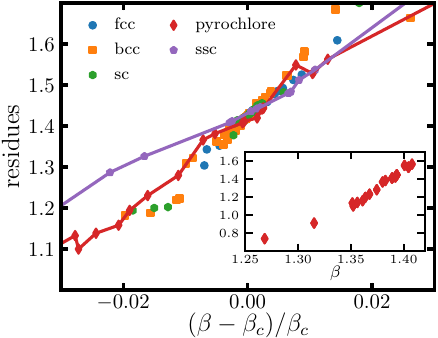}
\caption{Poles and residues from the Dlog Pad\'e method on $\overline{\chi}(\beta)$ for the ferromagnetic case on the \textit{fcc}, \textit{bcc}, \textit{sc}, \textit{ssc} and pyrochlore lattices as a function of $(\beta-\beta_c)/\beta_c$. The inset shows the poles and residues for the pyrochlore lattice.}
\vspace{-0.5cm}
\label{fig3}
\end{center}
\end{figure}

So far, $\beta_c$ and $\gamma$ have been obtained from standard methods, albeit with more orders. Alternatively, we can use our knowledge of $\gamma$ to get $\beta_c$ (see Fig. \ref{fig3}). In the Dlog Pad\'e method, each Pad\'e approximant provides a singularity at a given $\beta$, and its residue gives the value of the critical exponent. In practice, these couples of $\beta$ and $gamma$ are not randomly scattered and, instead, fall over a monotonic increasing function. Larger values of $\beta$ are accompanied by larger values of $\gamma$. In the end, for all lattices, the residues from Fig.~\ref{fig1} plotted versus their poles $\beta$ from Fig.~\ref{fig2} can be plotted, and the intersection with $\gamma\simeq 1.4$ gives an approximation of $\beta_c$. Thus, with a correct choice of $\beta_c$, the residues versus $(\beta-\beta_c)/\beta_c$ should collapse on lines crossing at the universal $\gamma$ for all lattices. Surprisingly, the lines of the \textit{fcc}, \textit{bcc}, \textit{sc}, and pyrochlore lattices present similar slopes, whereas it is smaller for \textit{ssc}. The \textit{ssc} line gives residues at $\gamma$ for $\beta_c = 4.20(5)$ (see Fig. \ref{fig3}). The behavior of these lines could help to determine critical values when the points do not accumulate near a single point $(\beta_c,\gamma)$, as in the \textit{ssc} lattice.

Another alternative is to use the Diagonal Dlog Pad\'e method as presented in Ref.~\cite{Kuzmin20}. In this method, the Pad\'e approximants of the inverse logarithmic derivative of $\chi$ are calculated, and only those with the same order in the denominator and numerator (the diagonal ones) are taken into account. Estimates of $\beta_c$ are then obtained from the least positive root of the numerator (unless it is also a root of the denominator). These values converge when the order $n$ increases, as illustrated for the pyrochlore and \textit{ssc} lattices in Table~\ref{tab1}. In the case of the pyrochlore, we obtain $\beta_c = 1.39(1)$ in agreement with our previous results. However, for the \textit{ssc} lattice we get a more precise estimate $\beta_c = 4.20(1)$.

\begin{table}[!t]
\centering
\begin{tabular}{c|ccccccc}
\hline
$n$ &   8 &  10 & 12 & 14 & 16 & 18 & 20 \\
\hline
 pyrochlore  &  1.314 &  1.408 &  1.380 &  1.394 & 1.394  &    &   \\
 \textit{ssc} &  5.043 & 4.343 & 4.359 & 4.351 & 4.206 & 4.209 & 4.202 \\
\end{tabular}
\caption{Diagonal Dlog Pad\'e results for $\beta_c$ for the ferromagnetic case on the \textit{ssc} and pyrochlore lattices.}
\label{tab1}
\end{table}

\vspace{-0.3cm}
\subsection{Dlog Pad\'e method applied to $c_v(\beta)$}
\label{sec:Dlogres_cv}
\vspace{-0.3cm}
	
So far, $\beta_c$ and the critical exponent $\gamma$ have been determined using the ferromagnetic susceptibility, which presents a strong singularity. The problem with relying on $\chi$ is that it depends on the order parameter, which is not generally known. Even in cases where it is known, like for antiferromagnetic models ($J>0$) presenting a phase transition (bipartite lattices), new HTSEs for the antiferromagnetic staggered susceptibility $\chi_{\rm AF}$ have to be calculated in order to see a strong singularity~\cite{Oitmaa04}, which is computationally expensive. On the other hand, other thermodynamic functions such as the specific heat $c_v(\beta)$ and the entropy $s(\beta)$ are always indicative of a phase transition, and ferro- and antiferromagnetic models are connected by the transformation $\beta \rightarrow -\beta$. So the advantage is that no new HTSEs have to be calculated. However, there is also a disadvantage. In these functions, the ferro- and antiferromagnetic singularities coexist on the HTSE and are always present on the positive and negative $\beta$ axis. Keeping this in mind, we now try to characterize a phase transition in the universality class of the Heisenberg 3D model, but without knowing the order parameter (i.e. $\chi$): for this, we now focus on the $c_v(\beta)$ function. 

Since $c_v$ behaves as $B - A \left(\beta_c - \beta \right)^{-\alpha}$ with $-1<\alpha<0$ close to $\beta_c$, the Dlog Pad\'e method cannot be used directly (the logarithmic derivative of $c_v$ does not have a simple pole at $\beta_c$). However, the Dlog Pad\'e method can be used on $c_v -B$. Doing so provides a good number of poles near the accepted $\beta_c$. As $B$ is a priori unknown, we can select its value such that we get the highest quality of results and then deduce $\alpha$. The issue is that there is a wide range of $B$ values that give high-quality results. However, this method allows us to obtain a well-defined dependency between the height of the peak $B$ and the critical exponent $\alpha$. The resulting $B(\alpha)$ is displayed in Figs.~\ref{fig7} and \ref{fig8}, together with our interpolation method results.

\vspace{-0.3cm}
\subsection{Dlog Pad\'e method applied to $\overline{\chi}(e)$}
\label{sec:Dlogres_chie}
\vspace{-0.3cm}

Finally, we can also study the singularity in $\overline{\chi}(e)$, where $e$ is the energy per site. To determine the type of singularity of this function at the transition, occurring at the critical energy $e_c$, we start from the singularity in $c_v$, which can be re-written as 
\begin{equation}
c_v^s(\beta) = B - \tilde{A} \Delta T^{-\alpha} 
\end{equation}
where $\tilde{A} = A/T_c^{-2\alpha}$ and $\Delta T = T - T_c$. By integration, we get that close to the critical point
\begin{equation}
\Delta e = B \Delta T + \frac{\tilde{A}}{1-\alpha} \Delta T^{1-\alpha} + o(\Delta T^{1-\alpha})
\end{equation}
where $\Delta e = e - e_c$. To get information on the singularity in $\overline\chi(e)$ we need the inverse $\Delta T (\Delta e)$. Since $\alpha$ is negative, the leading order is in $\Delta T$. Then
\begin{equation}
\Delta T = \frac{\Delta e}{B} - \frac{\tilde{A}}{(1-\alpha)B^{2-\alpha}} \Delta e^{1-\alpha} + o(\Delta e^{1-\alpha})
\end{equation}
Keeping only the leading order and knowing that $\overline \chi^s(T) \propto \Delta T^{-\gamma}$ leads to the simple result that $\overline{\chi}^s(e) \propto \Delta e^{-\gamma}$, and the Dlog Pad\'e method should give $e_c$ as pole and $\gamma$ as residue. However, taking into account that $\alpha$ is between $-0.1$ and $-0.2$, the second leading term has similar order compared to the leading term. The quotient between both terms depends on $\Delta e^{-\alpha}$. This corresponds to a cusp-like singularity that reaches 0 only at the critical energy $\Delta e = 0$. For example, using typical values of $A$, $B$, and $T_c$ on the fcc lattice, this quotient is about 1 when $\Delta e = 1$ and about 0.25 when $\Delta e = 0.0001$. In conclusion, only singularities at $\Delta e = e - e_c = 0$ exist, but the simple pole assumption is valid only infinitesimally close to the critical point and therefore the HTSEs are not able to represent it accurately. Thus, the Dlog Pad\'e method can be used to obtain values for $e_c$ from the poles, but the residues cannot capture the values of $\gamma$ or $\alpha$.

\begin{figure}[!t]
    \begin{center}
        \includegraphics*[width=0.4\textwidth]{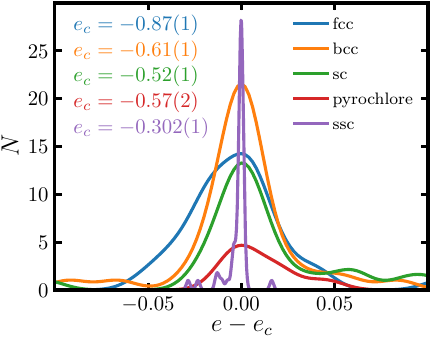}
        \caption{Density of poles from the Dlog Pad\'e method on $\overline{\chi}(e)$ for the ferromagnetic model. Results are shown as a function of $e - e_c$, where $e_c$ is the value at which there is a peak. For each lattice, poles from HTSEs at orders from $n-3$ to $n$ are used, where $n$ is given in Tab.~\ref{tab:HTSEorders}.}
\vspace{-0.5cm}
        \label{fig4}
    \end{center}
\end{figure}

From the HTSEs of $\beta f(\beta)$ and of $\overline{\chi}(\beta)$ at order $n$, we obtain the series of $\overline{\chi}(e)$ at order $n-1$ (because the series of $e(\beta)$ are of order $n-1$). Then we use the Dlog Pad\'e method on $\overline{\chi}(e)$ and obtain the critical energies for all lattices in the ferromagnetic case (see Fig. \ref{fig4}). We use Eq.~\eqref{eqN} with $e$ (instead of $\beta$) and $\sigma = 0.01$ for the \textit{fcc}, \textit{bcc}, \textit{sc} and pyrochlore lattices, and $\sigma = 0.001$ for the \textit{ssc} lattice. We find $e_c = -0.87(1)$ for the \textit{fcc} ($\sim 58\%$ of the ground-state energy), $e_c = -0.61(1)$ for the \textit{bcc} ($\sim 62\%$), $e_c = -0.52(1)$ for the \textit{sc} ($\sim 70\%$), $e_c=-0.57(1)$ for the pyrochlore ($\sim 76\%$), and $e_c = -0.302(1)$ for the \textit{ssc} ($\sim 81\%$). Contrary to the $\overline{\chi}(\beta)$ case, the method on $\overline{\chi}(e)$ works notably better for the \textit{ssc} lattice than for the rest. The pyrochlore lattice has a noticeably lesser amount of poles around $e_c$. The reason is again that this lattice is not bipartite, and the antiferromagnetic solutions on the positive $e$ axis are frustrated. This leads to a large number of poles appearing in the positive $e$ axis at values $e^* < |e_c|$. Finally, the residues are different for all lattices, indicating a dependency on non-universal quantities such as $A$, $B$, and $T_c$. As was expected from our previous analysis, $\alpha$ or $\gamma$ cannot be extracted.

\begin{figure}[!t]
\begin{center}
\includegraphics*[width=0.45\textwidth]{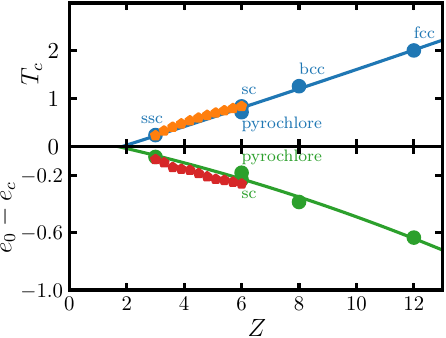}
\caption{Critical temperature $T_c$ (top panel) and the difference between the ground-state and critical energies $e_0-e_c$ (bottom panel) as a function of the coordination number $Z$ for the ferromagnetic model on the \textit{fcc}, \textit{bcc}, \textit{sc}, \textit{ssc} and pyrochlore lattices obtained from the Dlog Pade method on $\overline{\chi}$. We also show the interpolation between the \textit{sc} and \textit{ssc} lattices by using an effective $Z$ (see main text).}
\vspace{-0.5cm}
\label{fig6}
\end{center}
\end{figure}

\vspace{-0.3cm}
\subsection{$Z=2$ limit}
\vspace{-0.3cm}

We summarize our results obtained with the Dlog Pad\'e method for \textit{fcc}, \textit{bcc}, \textit{sc}, pyrochlore, and \textit{ssc} lattices in Fig.~\ref{fig6}. We plot the critical temperature $T_c$ extracted from $\overline \chi(\beta)$, together with the difference between the critical energy $e_c$ extracted from $\overline \chi(e)$ and the ground state energy $e_0$ (known exactly for the ferromagnetic case), as a function of the coordination number $Z$. Both $T_c$ and $e_c$ show a linear behavior with respect to $Z$, and $Z=2$ is a critical point for the finite temperature transitions in 3D ferromagnets \cite{Kuzmin20}, corresponding to a one-dimensional chain, characterized by $T_c=0$ and $e_c=e_0$. To get more points, we define the \textit{sc-ssc} model, interpolating between \textit{sc} and \textit{ssc} lattices (for which we have the HTSEs up to order $n=13$), with two types of links on the cubic lattice:
\begin{equation}
\mathcal{H} =  J_1\sum_{\langle i j \rangle} \mathbf{S}_i \cdot \mathbf{S}_j +  J_1' \sum_{\langle i j \rangle'} \mathbf{S}_i \cdot \mathbf{S}_j  
\label{eq-sc-ssc}
\end{equation}
in such a way that when $J_1=J_1'=1$ we have the \textit{sc} lattice. When $J_1 = 1$ and $J_1'=0$ (or vice versa), the Hamiltonian becomes that of the \textit{ssc} lattice. We also know that the coordination number goes from $Z=3$ at $J_1'=0$ to $Z=6$ at $J_1'=1$, so we can define an effective coordination number $Z_\text{eff}(J_1')=3(J_1+J_1')$ such that $e_0$ is proportional to $Z_\text{eff}$. The discrepancies between pentagons and circles of Fig.~\ref{fig6} at $Z=3$ and 6 (more visible for the energies) are due to different HTSE orders (13 for the \textit{sc-ssc} lattice, and 17 and 20 for the \textit{sc} and \textit{ssc}, respectively). In addition, we also continued our calculations for the frustrated case $J_1'/J_1 < 0$ and found that $T_c$ vanishes for $\left(J_1'\right)_c = -0.15(3)J_1$.

\vspace{-0.3cm}
\subsection{Interpolation methods for $c_v(\beta)$}
\label{subseccv}
\vspace{-0.3cm}

\begin{figure}[!t]
\begin{center}
\includegraphics*[width=0.465\textwidth]{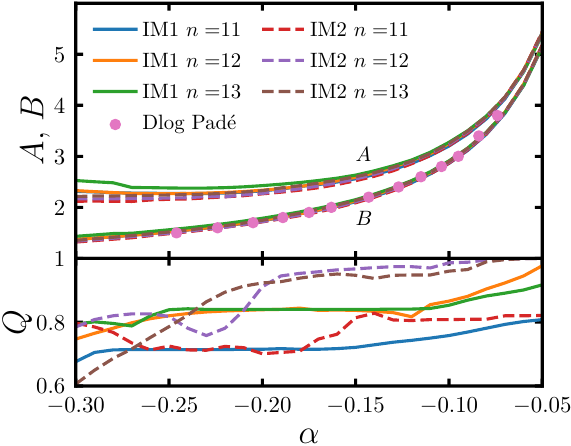}
\caption{Top panel: values of the singularity parameters $A$ and $B$ as a function of $\alpha$ for the ferromagnetic case on the \textit{fcc} lattice obtained from the two interpolation methods IM1 and IM2. We show the results for the three highest orders of the HTSE of $c_v$. The pink dots are the results obtained with the Dlog Pad\'e method on $c_v(\beta) - B$. Bottom panel: the quality $Q$ (see Eq.~\eqref{eq:Q}).}
\vspace{-0.5cm}
\label{fig7}
\end{center}
\end{figure}

The interpolation methods IM1 and IM2 presented in Sec.~\ref{sec:interpolation} for $c_v(\beta)$ have a three dimensional parameter space: $(\beta_c, A, \alpha)$ and $(\beta_c, B, \alpha)$ for IM1 and IM2, respectively. It is one more than the similar method used for logarithmic divergencies \cite{Gonzalez21}, which does not have to determine a critical exponent. Exploring the whole parameter space is thus time-consuming, so it is convenient to rely on other methods to narrow down some of the dimensions. In this sense, the Dlog Pad\'e method on $\overline \chi$ studied in Sec.~\ref{sec:Dlogres_chi} provides accurate values for the inverse critical temperature, $\beta_c$. Thus, we leave this parameter fixed. Regarding the value of $\alpha$, we know that the renormalization group value is $-0.122(10)$ \cite{Guida98}, while indirect estimations from HTSEs throw out values up to $-0.200(15)$ \cite{Kuzmin19}. So, for this parameter, we will search in the range $[-0.05,-0.30]$ at 0.01 intervals. For each value of $\alpha$ we search for the best value of $A$ or $B$ (for IM1 or IM2) in a range from 0.1 to 9, at 0.002 intervals. This step must be small since the peaks in $Q(A)$ for a given $\alpha$ and $\beta_c$ tend to be very narrow.

Fig.~\ref{fig7} shows the $A$ and $B$ values depending on the choice of $\alpha$, obtained through IM1 and IM2, for the \textit{fcc} lattice, together with the Dlog Pad\'e method results on $c_v - B$, discussed in Sec.~\ref{sec:Dlogres_cv}. The results for $A$ and $B$ show a good convergence with the HTSE order (especially at higher values of $\alpha$). Furthermore, there is a good agreement between all three methods. Let us recall that IM1 and IM2 remove the singularity by subtracting and dividing, respectively, such that for IM1, $A$ is a fitting parameter and $B$ is a byproduct. For IM2 it is the other way around. The quality $Q$ takes high values throughout all the $\alpha$ range: over $80\%$ of Pad\'e approximants coincide past $\beta_c$ (up to $(1+\delta)\beta_c$). Even though there is a tendency to higher $Q$ values for $\alpha$ closer to 0, it is not possible to pick one good value for $\alpha$, even choosing more restricting values for the $Q$-parameters, $\delta$, and $\epsilon$.

\begin{figure}[!t]
\begin{center}
\includegraphics*[width=0.45\textwidth]{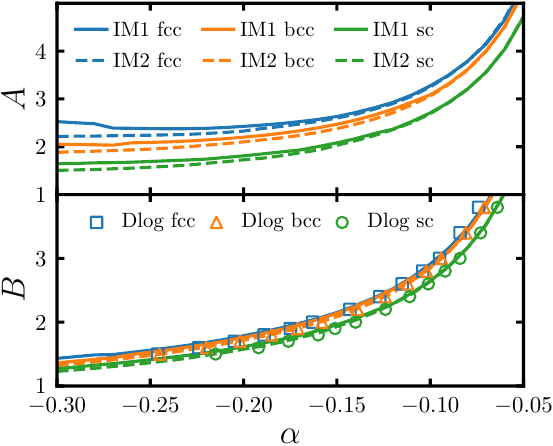}
\caption{Values of the non-universal parameters $A$ and $B$ as a function of $\alpha$ for the ferromagnetic model on the \textit{fcc}, \textit{bcc} and \textit{sc} lattices; using only the highest order in the HTSEs. Results obtained with IM1 and IM2 are shown with full and dashed lines, while symbols correspond to the Dlog Pad\'e method on $c_v(\beta) - B$.}
\vspace{-0.5cm}
\label{fig8}
\end{center}
\end{figure}

For the \textit{bcc} lattice, $Q$ is between 0.7 to 0.9, whereas on the \textit{sc} lattice, goes from 0.5 to 0.7 as $\alpha$ gets closer to zero. However, having half of the Pad\'e approximants down to $T_c$ is still a very good solution, since none of them are the same at that point when taking the raw HTSE. Also, these lattices show a convergence of the $A$ and $B$ values with the HTSE order $n$ that is similar to the \textit{fcc} lattice, using IM1 and IM2. $A$ and $B$ values for the highest HSE order are given in Fig.~\ref{fig8} for the \textit{fcc},\textit{bcc} and \textit{sc} lattices. For all lattices, IM1 and IM2 give similar results, especially in the case of $B$. Differences only show up for $A$ at values of $\alpha$ far from 0. We obtained higher values for $A$ and $B$ than those in the literature using HTSEs \cite{Baker67, Oitmaa96}. However, the latter were calculated by fitting the Pad\'e approximants of the raw $c_v$-HTSEs with the critical behavior from Eq.~\ref{eq:cvsing} in a small range $\beta < 0.96\ \beta_c$.

Another interesting feature is that the values of $B$ (the peak height) are very similar for the three lattices, while the values of $A$ seem to change slowly as the coordination number $Z$ changes. A similar thing happens with the parameters of the singularities on Ising models on 2D, where the values are very similar but not universal in different lattices \cite{Gonzalez21}. For the pyrochlore lattice, $Q$ takes lower values, between 0.3 and 0.4. However, we can still extract values of $A$ and $B$. They are smaller than in the \textit{sc} lattice, even though both have the same coordination number $Z$. Taking these four lattices, $A$ and $B$ decrease as $T_c$ decreases. When $Z=2$ there is no finite-temperature phase transition, so it might be interesting to see how this limit is reached in terms of $A$ and $B$. Finally, for the \textit{ssc} lattice, no clear peak can be determined.

Even though it is not possible to determine the critical exponent $\alpha$ with these methods, we obtain well-defined functions for $A(\alpha)$ and $B(\alpha)$. Thus, we can reconstruct $c_v$ above $T_c$ for a supposed value of $\alpha$. Using the reconstructed $c_v$ to calculate the critical energy $e_c$ by integration, which depends on the parameters, one could attempt to determine $\alpha$ by comparing with the Dlog Pad\'e results. For the \textit{fcc} lattice, the values of $e_c$ go from $-0.860$($-0.857$) to $-0.867$ ($-0.867$) for IM1 (IM2) as $\alpha$ changes from $-0.3$ to $-0.05$. Again, higher values of $\alpha$ show a better agreement between methods. These values, summed up as $e_c=-0.862(5)$, are in agreement with the Dlog Pad\'e estimation from the previous section ($e_c = -0.87(1)$). Unluckily, the Dlog Pad\'e method does not offer sufficiently precise values of the energy and the function $e_c(\alpha)$ obtained by integration changes very little. So that in the end, it is not possible to use this extra information to determine the value of $\alpha$, which remains elusive. The same happens for all the remaining lattices. For the \textit{bcc} lattice we get $e_c=-0.607(3)$, in agreement with our Dlog Pad\'e result $e_c = -0.61(1)$. For the \textit{sc} lattice $e_c=-0.511(2)$, in agreement with $e_c=-0.52(1)$ from the Dlog Pad\'e method. Finally, for the pyrochlore we get $e_c=-0.578(3)$, in agreement with $e_c=-0.57(2)$ obtained from the Dlog Pad\'e method. 

\begin{figure}[!t]
\begin{center}
\includegraphics*[width=0.45\textwidth]{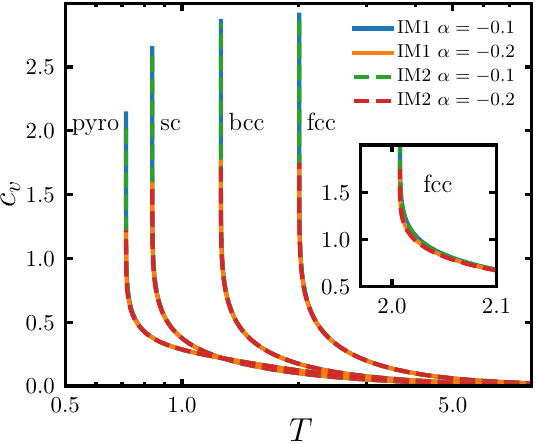}
\caption{Reconstructed $c_v$ from the two interpolation methods for the ferromagnetic case on the \textit{fcc}, \textit{bcc}, \textit{sc}, and pyrochlore lattices. We show the results for two different values of $\alpha$. The inset shows a zoom for the \textit{fcc} lattice close to the critical temperature.}
\vspace{-0.5cm}
\label{fig9}
\end{center}
\end{figure}

The critical entropy can also be obtained by integration. Let us start with the \textit{sc} lattice, for which we obtain $s_c=0.402(1)$, where $s_c=0.401$ for $\alpha =-0.1$ and $s_c=0.403$ for $\alpha=-0.2$ with both interpolation methods. In this case, it is possible to benchmark with QMC results, which give $s_c= 0.401(5)$ \cite{Wessel10}. All of our $s_c$ values are in agreement with QMC while showing a slightly better precision. Also, it is not possible to decide on $\alpha$ from this calculation. For the \textit{bcc} lattice, all of our results are within $s_c=0.435(1)$. For the \textit{fcc} lattice we get $s_c = 0.443(1)$. Finally, for the pyrochlore, we obtain $s_c=0.353(2)$. All in all, we can see that the critical entropy decreases with the coordination number, as does the critical temperature. However, contrary to the results shown in Fig.~\ref{fig6} for the critical temperature and $e_0-e_c$, the critical entropy $s_c$ does not show a linear behavior towards 0 at $Z=2$.

\begin{figure}[!t]
\begin{center}
\includegraphics*[width=0.47\textwidth]{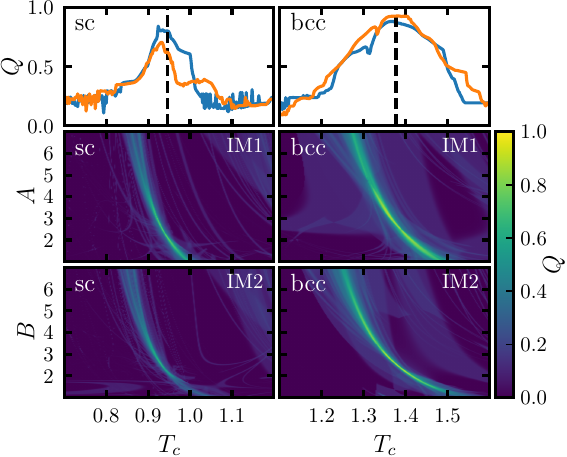}
\caption{Interpolation method results for $c_v(\beta)$ for the antiferromagnetic model on the \textit{sc} (left) and \textit{bcc} (right) lattices. The quality of results $Q$ is shown in color scale in a region of the parameter space defined by $T_c$ and $A$ or $B$. The critical exponent $\alpha$ is set to $-0.12$. The quality $Q$ as a function of $T_c$ is shown in the top panels, in blue and orange for IM1 and IM2, respectively. Dashed black lines indicate QMC results \cite{Sandvik98, Ghosh19b}.}
\vspace{-0.5cm}
\label{fig12}
\end{center}
\end{figure}

We show in Fig.~\ref{fig9} the reconstructed $c_v$ for the four lattices (\textit{fcc}, \textit{bcc}, \textit{sc} and pyrochlore) using the best values of $A$ and $B$ at the accepted $\beta_c$ for two limiting values of $\alpha$, $-0.2$ and $-0.1$. Both methods give the same curves, and the differences between the two values of $\alpha$ can only be seen very close to the corresponding critical points (see inset for \textit{fcc}) through a very different value of the peak height $B$, as can be seen in the previous figures of $B(\alpha)$. However, this issue only exists at exactly the critical temperature, so it does not affect the comparison with experimental results since the sharp theoretical peaks with divergent slopes at the critical point cannot be captured by experiments in real compounds \cite{Lederman74, Kornblit75, Haeiwa88, Khan12, Berry23}. To sum up, we have a good precision for every temperature above the critical temperature $T_c$ obtained from finite high-temperature series expansions. Thus, this method extrapolates the specific heat from HTSE down to almost the critical temperature for the phase transitions of several ferromagnetic Heisenberg models.

As we mentioned earlier, the advantage of using $c_v$ instead of $\chi$ resides in the possibility of studying both ferro- and antiferromagnetic models with the same HTSE. In Fig.~\ref{fig12} we show results obtained for the antiferromagnetic \textit{sc} and \textit{bcc} lattices. For these calculations, since one wants to avoid using data for $\overline{\chi}(\beta)$, we calculated $T_c$ by fixing $\alpha=-0.12$, close to the field theory result. However, the critical temperature is not very sensitive to this constraint, at least within the range of values reported in the literature. In the top panels of Fig.~\ref{fig12} we show the values of $Q$ as a function of $T_c$ obtained from the color plots. For both lattices, there is a well-defined region of parameters with high $Q$, meaning that almost all Pad\'e approximants coincide down to the critical temperature. Also, these regions of high $Q$ agree well with QMC results, indicated by the dashed black lines  \cite{Sandvik98, Ghosh19b}. For the \textit{sc} lattice we get $T_c = 0.93(2)$, in agreement with the QMC result $T_c = 0.946(1)$ \cite{Sandvik98}. For the \textit{bcc} lattice we get 1.38(4), in agreement with the QMC result $T_c = 1.377(2)$  \cite{Ghosh19b}. We can also take the best values from each interpolation method to reconstruct $c_v$ for the antiferromagnetic cases and calculate the critical energies and entropies. For the \textit{sc} lattice, we get $e_c = -0.66(1)$ and $s_c = 0.331(5)$. The latter is slightly lower than the QMC result, $s_c=0.341(5)$ \cite{Wessel10}, but still fairly close. For the \textit{bcc} lattice we get $e_c = -0.73(1)$ and $s_c = 0.393(5)$.

\vspace{-0.3cm}
\subsection{Interpolation method for $\chi(\beta)$}
\vspace{-0.3cm}

\begin{figure}[!t]
\begin{center}
\includegraphics*[width=0.47\textwidth]{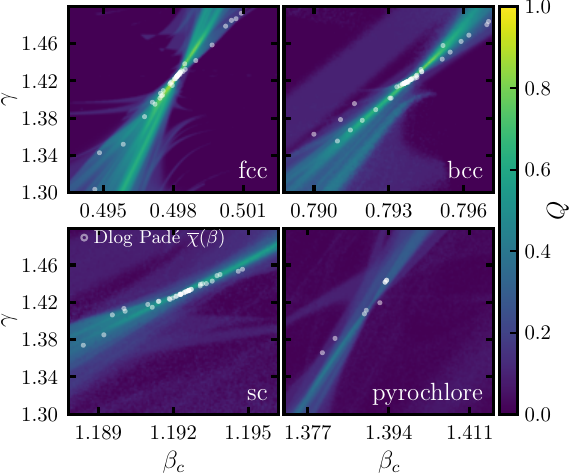}
\caption{Interpolation method IM2 results for $\overline{\chi}(\beta)$ for the ferromagnetic case on the \textit{fcc}, \textit{bcc}, \textit{sc} and pyrochlore lattices. The quality of results $Q$ is shown in color scale in a region of the parameter space $\{\beta_c, \gamma\}$ close to the best values. White circles indicate the poles and residues from the Dlog Pad\'e method.}
\vspace{-0.5cm}
\label{fig10}
\end{center}
\end{figure}

Finally, we apply the interpolation method IM2 to the ferromagnetic $\overline{\chi}(\beta)$ (as explained in Sec.~\ref{sec:interpolation2} for divergent singularities). The parameter space is two-dimensional $\{\beta_c, \gamma\}$, and the region of high-quality values is narrow so both parameters have to be calculated using a fine mesh.

Fig.~\ref{fig10} shows the quality $Q$ as a function of $\beta_c$ and $\gamma$ for the \textit{fcc}, \textit{bcc}, \textit{sc}, and pyrochlore lattices. We also show in white circles the results from the poles and residues obtained from the Dlog Pad\'e method. For all lattices, the poles and residues are concentrated around the large $Q$ region from IM2, and reciprocally, the higher $Q$ values are obtained close to the line of poles and residues from the Dlog Pad\'e method. This illustrates a close connection between both methods. Specifically, the \textit{fcc} lattice presents $Q=1.00$ around $\beta_c = 0.4981(2)$ and $\gamma = 1.422(2)$, the \textit{bcc} lattice presents $Q=0.93$ for $\beta_c=0.7938(2)$ and $\gamma = 1.420(3)$, the \textit{sc} lattice presents $Q=0.64$ for $\beta_c = 1.1935(10)$ and $\gamma = 1.44(1)$. All of these are in agreement with the Dlog Pad\'e results. For the pyrochlore and \textit{ssc} (not shown) lattices, the results are not so clear. The pyrochlore lattice has a large cloud of values $Q=0.41(1)$ along a well-defined line around $\beta_c = 1.382(5)$ and $\gamma = 1.36(3)$. However, the diagonal Dlog Pad\'e results lie closer to the endpoint of this cloud. For the \textit{ssc} lattice, there are just a few points around $Q=0.23(2)$ with $\beta_c = 4.20(2)$ and $\gamma = 1.35(2)$, but the quality is too low to consider them reliable.

\vspace{-0.3cm}
\section{Conclusions and perspectives}
\label{secCon}
\vspace{-0.3cm}

We have studied the finite-temperature phase transition that occurs in ferromagnetic quantum Heisenberg models on 3D lattices by using several methods derived from the HTSEs. We used the standard Dlog Pad\'e method and estimated $\beta_c$ and $\gamma$ of the \textit{fcc}, \textit{bcc}, \textit{sc}, pyrochlore and \textit{ssc} lattices. For some of them, results are given for larger orders of HTSEs than in the previous works. However, the discrepancy in $\gamma$ between field theory's renormalization group for the classical case and the HTSEs results for the quantum case is still present. Also, no convergence towards the classical values is observed, and this remains an open question. We have also explored possible extensions of these methods. While standard calculations involve $\chi(\beta)$, we have obtained the critical energy $e_c$ using $\chi(e)$. We also used the Dlog Pad\'e method on $c_v(\beta)-B$ (with $B=c_v(\beta_c)$) to obtain $B(\alpha)$.

\begin{table}[t!]
\centering
\begin{tabular}{llllll}
\hline
\hline
& \multicolumn{2}{c}{$\beta_c$} & \multicolumn{2}{c}{$e_c$} & \multicolumn{1}{c}{$s_c$}\\
\hline
\multicolumn{1}{l}{Lattice} & \multicolumn{1}{c}{DLP} & \multicolumn{1}{c}{IM} & \multicolumn{1}{c}{DLP} & \multicolumn{1}{c}{IM}& \multicolumn{1}{c}{IM}\\
\hline
fcc & 0.4982(2) & 0.4981(2) &  -0.87(1) & -0.862(5) & 0.443(1)\\
bcc & 0.7937(2) & 0.7938(2) &  -0.61(1) & -0.607(3) & 0.435(1)\\
sc & 1.1926(2) & 1.1935(10) &  -0.52(1) & -0.511(2) & 0.402(1) \\
pyro & 1.39(1) & 1.382(5) & -0.57(2) & -0.578(3) & 0.353(2) \\
ssc & 4.20(1) & 4.20(2) & -0.302(1) & & \\
\hline
\hline
\end{tabular}
\caption{Summary of results for $\beta_c$, $e_c$, and $s_c$ obtained in this article for the ferromagnetic Heisenberg model on the \textit{fcc}, \textit{bcc}, \textit{sc}, pyrochlore (pyro), and \textit{ssc} lattices. DLP stands for Dlog Pad\'e method and IM for the interpolation method.}
\label{tab3}
\end{table}

Then we presented interpolation methods to obtain $c_v(T)$ and $\chi(T)$ for $T>T_c$. These methods are efficient for the \textit{fcc}, \textit{bcc}, and \textit{sc} lattices, but less for the pyrochlore and \textit{ssc} lattices. For $c_v(T)$, we are not able to get a precise value of the critical exponent $\alpha$, but the methods provide accurate relationships between the three important parameters at the singularity, $A$, $B$, and $\alpha$. Thus, if any of them is known, the other two can be deduced. We have also shown that the interpolated $c_v(T)$ has a very small dependence on $\alpha$ as soon as $T$ is slightly above $T_c$, the main difference being in the value of the peak at $T_c$. This allowed us to obtain accurate results of the critical energy $e_c$ and critical entropy $s_c$ for most lattices studied. The main advantage of the method is that it can be directly applied to antiferromagnetic cases without the need of calculating a new HTSE. To show this, we applied both interpolation methods to the \textit{bcc} and \textit{sc} lattices, and obtained results in agreement with QMC for $T_c$ and $s_c$. Finally, we also applied the interpolation method to obtain $\chi(T)$ above $T_c$, obtaining reliable values of $T_c$ and $\gamma$.

\begin{table}[t!]
\centering
\begin{tabular}{lllllll}
\hline
\hline
& \multicolumn{2}{c}{$\gamma$} & \multicolumn{2}{c}{$\alpha=-0.1$}  & \multicolumn{2}{c}{$\alpha=-0.2$} \\
\hline
\multicolumn{1}{l}{Lattice} & \multicolumn{1}{c}{DLP} & \multicolumn{1}{c}{IM} & \multicolumn{1}{c}{$A$} & \multicolumn{1}{c}{$B$}& \multicolumn{1}{c}{$A$} & \multicolumn{1}{c}{$B$}\\
\hline
fcc & 1.426(2) & 1.422(2) & 3.25(5) & 2.90(5) & 2.41(5) & 1.74(3)\\
bcc & 1.419(2) & 1.420(3) & 3.10(5) & 2.85(5) & 2.17(6) & 1.71(2)\\
sc &  1.433(3) & 1.44(1)  & 2.71(3) & 2.62(5) & 1.80(3) & 1.56(3)\\
pyro &         & 1.36(3)  & 2.15(15)& 2.1(1)  & 1.33(3) & 1.22(3)\\
ssc &          & 1.35(2)  &         &         &         &\\
\hline
\hline
\end{tabular}
\caption{Summary of results for $\gamma$, $A$, and $B$ obtained in this article for the ferromagnetic Heisenberg model on the \textit{fcc}, \textit{bcc}, \textit{sc}, pyrochlore (pyro), and \textit{ssc} lattices. DLP stands for Dlog Pad\'e method and IM for the interpolation method.}
\label{tab4}
\end{table}

In conclusion, we have probed several different methods based on HTSEs to study finite-temperature phase transitions. These methods allowed us to obtain accurately several quantities related to the critical points, such as critical exponents, critical temperatures, and parameters related to the singularities. As a summary, we present the main numerical results in Tables~\ref{tab3} and \ref{tab4}, where DLP stands for the Dlog Pad\'e results and IM stands for the interpolation method. The results shown for $A$ are obtained with IM, while the ones for $B$ cover both IM and Dlog Pad\'e results. It is important to note that even if we have approached only the ferro and antiferromagnetic Heisenberg models on the most common lattices without frustration, these methods are suitable for studying any kind of system with the same type of phase transitions.

\vspace{-0.3cm}

\section*{Acknowledgments}

\vspace{-0.3cm}

The authors would like to thank Andrey Zabolotskiy for pointing out towards skipped orders at the end of Ref.~\cite{Oitmaa96}, and for including our new coefficients into the Online Encyclopedia of Integer Sequences. This work was supported by the French Agence Nationale de la Recherche under Grant No. ANR-18-CE30-0022-04 LINK

\appendix

\vspace{-0.3cm}

\section{HTSEs for the $S=1/2$ models}
\label{ap1}

\vspace{-0.3cm}

In Tables \cref{tab:coefs1,tab:coefs2,tab:coefs3,tab:coefs4} we present the complete list of coefficients for the HTSEs of $\beta f$ and $\overline{\chi}$, where the new ones are in bold numbers. These are written in terms of Eq.~\ref{eq-defhtse} where $n_u=1$ for the \textit{fcc}, \textit{bcc}, and \textit{sc} lattices; $n_u=4$ for the \textit{ssc} and pyrochlore lattices.

\begin{table*}[t!]
\centering
\begin{tabular}{crrr}
\hline
\hline
\multicolumn{1}{c}{$a_n$} & \multicolumn{1}{c}{fcc} & \multicolumn{1}{c}{bcc} & \multicolumn{1}{c}{sc}\\
\hline
1 & $0$ & $0$ & $0$ \\
2 & $18$ & $12$ & $9$ \\
3 & $-108$ & $24$ & $18$ \\
4 & $180$ & $168$ & $-162$ \\
5 & $5040$ & $-1440$ & $-2520$ \\
6 & $162000$ & $24480$ & $33192$ \\
7 & $-14565600$ & $297024$ & $1019088$ \\
8 & $563253408$ & $28017216$ & $-7804944$ \\
9 & $-17544639744$ & $533681664$ & $-723961728$ \\
10 & $750412309248$ & $41156316672$ & $2596523904$ \\
11 & $-56646776913408$ & $503287538688$ & $856142090496$ \\
12 & $4973976625190400$ & $53001415916544$ & $6383648984832$ \\
13 & $\boldsymbol{-421817449494804480}$ & $1839416689004544$ & $-1356696930401280$ \\
14 &                        & $246102905022713856$ & $-27667884260938752$ \\
15 &                         & $\boldsymbol{9001661201883684864}$ & $\boldsymbol{2908030732698175488}$ \\
16 &                          &                       & $\boldsymbol{122264703581556307968}$\\
17 &                           &                       & $\boldsymbol{-7238339805811283361792}$\\
\hline
\hline
\end{tabular}
\caption{Coefficients $a_n$ corresponding to $\beta f$ as defined in Eq.~\ref{eq-defhtse} for the \textit{fcc}, \textit{bcc}, and \textit{sc} lattices.}
\label{tab:coefs1}
\end{table*}

\vspace*{-1cm}

\begin{table*}[t!]
\centering
\begin{tabular}{crrr}
\hline
\hline
\multicolumn{1}{c}{$b_n$} & \multicolumn{1}{c}{fcc} & \multicolumn{1}{c}{bcc} & \multicolumn{1}{c}{sc}\\
\hline
1 & $-6$ & $-4$ & $-3$ \\
2 & $120$ & $48$ & $24$ \\
3 & $-3312$ & $-832$ & $-264$ \\
4 & $117360$ & $18400$ & $3960$ \\
5 & $-5104416$ & $-504384$ & $-74928$ \\
6 & $263405088$ & $16313280$ & $1584624$ \\
7 & $-15717292800$ & $-610699520$ & $-38523264$ \\
8 & $1063892512512$ & $25867292160$ & $1115604864$ \\
9 & $-80532234584064$ & $-1229543182336$ & $-35969253888$ \\
10 & $6741740335372800$ & $64541249655808$ & $1223162767104$ \\
11 & $-618536855295817728$ & $-3716345369001984$ & $-46443134693376$ \\
12 & $61718837768472705024$ & $232442811396567040$ & $1997899947119616$ \\
13 & $\boldsymbol{-6654017125619285385216}$ & $-15728092831910068224$ & $-90256082576916480$ \\
14 &                        & $1142407619109235630080$ & $4221503453720782848$ \\
15 &                         & $\boldsymbol{-88805626440393148956672}$ & $\boldsymbol{-220236945885669801984}$ \\
16 &                          &                       & $\boldsymbol{12562562473105938481152}$\\
17 &                           &                       & $\boldsymbol{-722105535259151290073088}$\\
\hline
\hline
\end{tabular}
\caption{Coefficients $b_n$ corresponding to $\overline{\chi}$ as defined in Eq.~\ref{eq-defhtse} for the \textit{fcc}, \textit{bcc}, and \textit{sc} lattices.}
\label{tab:coefs2}
\end{table*}

\begin{table*}[t!]
\centering
\begin{tabular}{crr}
\hline
\hline
\multicolumn{1}{c}{$a_n$} & \multicolumn{1}{c}{pyrochlore} & \multicolumn{1}{c}{ssc}\\
\hline
1 & $0$ & $\boldsymbol{0}$ \\
2 & $36$ & $\boldsymbol{18}$ \\
3 & $-72$ & $\boldsymbol{36}$ \\
4 & $-1656$ & $\boldsymbol{-324}$ \\
5 & $18720$ & $\boldsymbol{-3600}$ \\
6 & $340704$ & $\boldsymbol{20592}$ \\
7 & $-11342016$ & $\boldsymbol{788256}$ \\
8 & $-99460800$ & $\boldsymbol{-267552}$ \\
9 & $11144157696$ & $\boldsymbol{-292582656}$ \\
10 & $-43247015424$ & $\boldsymbol{-2338428672}$ \\
11 & $-15542133488640$ & $\boldsymbol{158857763328}$ \\
12 & $359762974166016$ & $\boldsymbol{3398565523968}$ \\
13 & $26719647518453760$ & $\boldsymbol{-111431579830272}$ \\
14 & $\boldsymbol{-1532961802218000384}$  & $\boldsymbol{-5116416515020800}$ \\
15 & $\boldsymbol{-44591351194841260032}$   & $\boldsymbol{83476825611595776}$ \\
16 & $\boldsymbol{6653104879154138357760}$   & $\boldsymbol{9038092645962510336}$                      \\
17 &  $\boldsymbol{-8491904138175731859456}$ & $\boldsymbol{-20724045060052942848}$                      \\
18 &                           & $\boldsymbol{-18839898190998133604352}$                      \\
19 &                           & $\boldsymbol{-253691725481243238334464}$                      \\
20 &                           & $\boldsymbol{45155117370822689756676096}$                      \\
\hline
\hline
\end{tabular}
\caption{Coefficients $a_n$ corresponding to $\beta f$ as defined in Eq.~\ref{eq-defhtse} for the pyrochlore and \textit{ssc} lattices.}
\label{tab:coefs3}
\end{table*}

\begin{table*}[t!]
\centering
\begin{tabular}{crr}
\hline
\hline
\multicolumn{1}{c}{$b_n$} & \multicolumn{1}{c}{pyrochlore} & \multicolumn{1}{c}{ssc}\\
\hline
1 & $-12$ & ${-6}$ \\
2 & $96$ & ${12}$ \\
3 & $-816$ & ${12}$ \\
4 & $8160$ & ${240}$ \\
5 & $-148992$ & ${-5136}$ \\
6 & $3879744$ & ${-40224}$ \\
7 & $-81019776$ & ${778464}$ \\
8 & $990764544$ & ${22859520}$ \\
9 & $-15165570048$ & ${-183876864}$ \\
10 & $1661765784576$ & ${-15637820928}$ \\
11 & $-97979429505024$ & ${-30648860160}$ \\
12 & $1761563239919616$ & ${14373541840896}$ \\
13 & $\boldsymbol{85410304429842432}$ & ${215523347675136}$ \\
14 & $\boldsymbol{-1996581576629084160}$  & ${-16345519886733312}$ \\
15 & $\boldsymbol{-507546664875986436096}$   & $\boldsymbol{-572943639086174208}$ \\
16 & $\boldsymbol{35052604281755859025920}$   & $\boldsymbol{20821759189681766400}$                      \\
17 &                           & $\boldsymbol{1524898473896350777344}$                      \\
18 &                           & $\boldsymbol{-22745675831893506785280}$                      \\
19 &                           & $\boldsymbol{-4446640583932914089459712}$                      \\
20 &                           & $\boldsymbol{-17616386456676250248806400}$                      \\
\hline
\hline
\end{tabular}
\caption{Coefficients $b_n$ corresponding to $\overline{\chi}$ as defined in Eq.~\ref{eq-defhtse} for the pyrochlore and \textit{ssc} lattices.}
\label{tab:coefs4}
\end{table*}

\clearpage

\bibliography{papers}

\end{document}